\documentclass[journal]{IEEEtran}

\usepackage{amsmath,amssymb}
\usepackage{graphicx}
\usepackage{tikz-cd}
\usepackage{multicol}

\usepackage{tikz}
\usepackage{pgf-pie}
\usepackage{pgfplots}
\pgfplotsset{compat=1.18}
\usepgfplotslibrary{colorbrewer}
\usetikzlibrary{shapes.geometric, arrows.meta, positioning, fit, backgrounds}
\usepackage{forest}
\useforestlibrary{edges}
\usepackage[numbers]{natbib}
\usepackage{booktabs}
\usepackage{multirow}
\usepackage{tabularx}
\usepackage{array}
\usepackage{stfloats}
\usepackage{xcolor}         
\usepackage{enumitem}       

\definecolor{mainrq}{RGB}{31, 73, 125}      
\definecolor{archblue}{RGB}{31, 119, 180}   
\definecolor{scopegreen}{RGB}{32, 128, 108} 
\definecolor{engpurple}{RGB}{102, 51, 153}  
\definecolor{qualrown}{RGB}{153, 76, 0}     
\definecolor{rq12}{RGB}{70, 150, 200}       
\definecolor{rq35}{RGB}{60, 160, 140}       
\definecolor{rq689}{RGB}{140, 90, 200}      
\definecolor{rq710}{RGB}{200, 110, 50}      

\usepackage[most]{tcolorbox}

\newtcolorbox{highlight}{
  colback=yellow!10,
  colframe=yellow!50!black,
  boxrule=0.5pt,
  arc=2pt,
  left=6pt,
  right=6pt,
  top=6pt,
  bottom=6pt
}

\usepackage[colorlinks=true, linkcolor=black, citecolor=black, urlcolor=black, bookmarks=false, pdfborder={0 0 0}]{hyperref}

\begin{document}

\title{LLM-based NPCs in Video Games: \\A Systematic Mapping Study}

\author{\fontsize{9}{11}\selectfont Dilan~Mian, Jeremy~Bradbury, Yann-Gael~Gu\'eh\'eneuc, Fabio~Petrillo, and~Cristiano~Politowski
\thanks{D. Mian, J. Bradbury, and C. Politowski are with Ontario Tech University, Oshawa, Canada (e-mail: dilan.mian@ontariotechu.ca; jeremy.bradbury@ontariotechu.ca; cristiano.politowski@ontariotechu.ca). Corresponding author: D. Mian.}
\thanks{Y.-G. Gu\'eh\'eneuc is with Concordia University, Montreal, Canada, (e-mail: yann-gael.gueheneuc@concordia.ca).}
\thanks{F. Petrillo is with \'Ecole de Technologie Sup\'erieure, Montreal, Canada, (e-mail: fabio.petrillo@etsmtl.ca).}
\thanks{\copyright~2026 IEEE. Personal use of this material is permitted. Permission from IEEE must be obtained for all other uses, in any current or future media, including reprinting/republishing this material for advertising or promotional purposes, creating new collective works, for resale or redistribution to servers or lists, or reuse of any copyrighted component of this work in other works.}%
\thanks{This article has been accepted for publication in IEEE Transactions on Games. DOI: 10.1109/TG.2026.3733302}}

\maketitle
\IEEEpubidadjcol
\vspace{-8pt}
\begin{abstract}
For decades, non-player characters (NPCs) have been driven by rule-based systems whose personalities and interactions are scripted before release. Large language models (LLMs) promise NPCs that converse openly, adapt to new situations, and behave in ways developers did not fully author, prompting a wave of research and commercial experimentation. Yet the field lacks a unified picture of what has been built, how well it works, and at what cost. We address this gap with a systematic mapping study of 377 papers from the academic literature, yielding 42 papers analysed against a main research question and ten sub-questions across four thematic axes. We provide the first quantitative characterisation of component adoption, latency, safety coverage, and reproducibility: short-term memory and world-state grounding are near-universal (60\% and 69\%), whereas long-term memory (31\%), reflection (38\%), and retrieval-augmented generation (14\%) remain rare, and response times span 2.6 to 15 seconds. Finally, we identify six structural barriers to commercial adoption and offer evidence-based recommendations for developers and researchers.
\end{abstract}


\section{Introduction}
\label{sec:intro}

Non-player characters (NPCs) shape player experience by delivering narrative, providing challenge, populating worlds, and mediating agency. Traditionally, explicitly scripted systems control their behaviour (dialogue trees, behaviour trees, and goal-oriented action planners), which brings consistency but is ultimately limited by what designers author in advance \cite{millington2009gameai}. 

Large Language models (LLMs) introduce the idea of NPCs that converse openly, adapt to new situations, and act in ways that are not pre-programmed. The growing accessibility of LLMs has led researchers and the industry to experiment with LLM-based NPCs, from video game mods\cite{llmpowerednpcs2024} like Skyrim to commercial platforms like InWorld AI.

However, this subject lacks a unified understanding of what has been built, how well it works, and at what cost. Existing studies use incompatible evaluation methods, customised environments, and undisclosed prompts, which hinder cross-comparison. No shared benchmarks for believability or dialogue coherence exist. Also, key issues such as latency, hallucination, and long-term consistency are rarely quantified, and few papers implement runtime guardrails.

We fill this gap by systematically mapping LLM-based NPC dialogue and behaviour across 377 records from ACM Digital Library, IEEE Xplore, and Scopus, yielding 42 papers for full analysis after applying inclusion, exclusion, and quality criteria. The review is structured around a main research question, \textit{``what LLM-based approaches exist for creating lifelike NPC dialogue and behaviour, and how are they evaluated?''}. To answer this, we derived \textit{ten sub-questions} across four thematic axes with 36 extracted variables.

This review makes three primary contributions. 
First, we present the first quantitative characterisation of component adoption, latency, safety coverage, and reproducibility across the corpus, showing that short-term memory and world-state grounding are near-universal (60\% and 69\%, respectively) while long-term memory (31\%), reflection (38\%), and RAG (14\%) remain minority practices, and that end-to-end response times range from 2.6 s to 15 seconds. 
Second, we identify and explain six structural barriers to commercial adoption (latency, quality-assurance breakdown, cost at player scale, multiplayer synchronisation, content-rating compliance, and localisation) that existing literature mentions in isolation, if at all. 
Finally, we offer evidence-based recommendations specifying which architecture patterns show the strongest signs of practical viability, what a minimum evaluation reporting standard should contain, and what the field's highest-priority open problems are.

The remainder of this paper is structured as follows.  
Section~\ref{sec:method} describes the systematic methodology. Section~\ref{sec:res} presents the results across all ten research questions. Section~\ref{sec:discussion} discusses the implications of the findings for developers and researchers. Section~\ref{sec:conclusion} concludes the paper.

\section{Methodology}
\label{sec:method}

\begin{figure}[htbp]
    \centering
    \includegraphics[width=\linewidth]{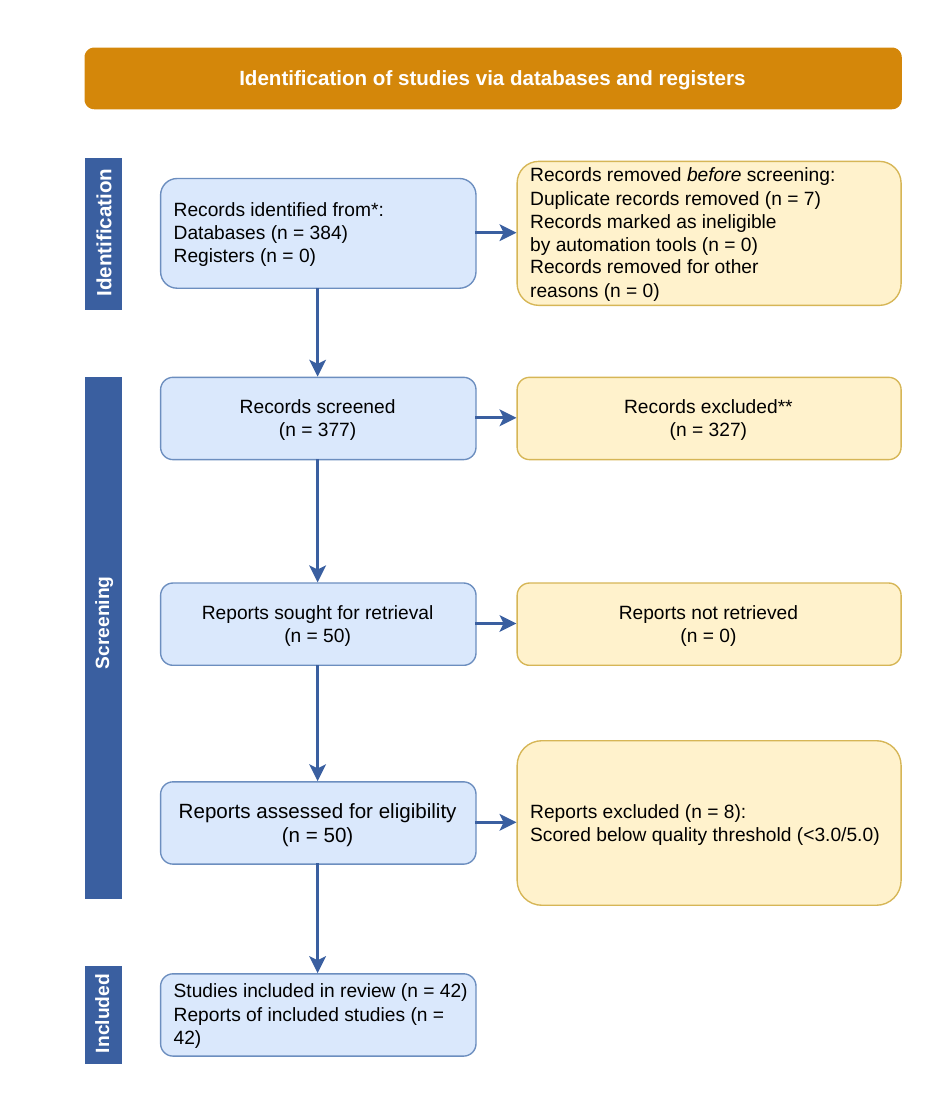}
    \caption{PRISMA flow diagram illustrating the systematic identification, screening, eligibility assessment, and final inclusion of studies across the review process.}
    \label{fig:prisma}
\end{figure}

We crafted a systematic mapping study following the guidelines of Sweetser~\cite{largelanguagemodels2025} to identify, evaluate, and synthesize research on LLM-based approaches to NPC behaviour and dialogue in interactive environments. We utilized the Prisma diagram in Figure~\ref{fig:prisma} to show the full screening flow of the selected corpus.
The process comprises the following five phases.

\subsection{Mapping Protocol and PICOC Framework}
\label{sec:method:picoc}

We grounded the mapping scope in a PICOC (Population, Intervention, Comparison, Outcome, Context) specification, which defines the boundaries of the primary studies' inclusion decisions~\cite{largelanguagemodels2025}. Table~\ref{tab:picoc} presents the five elements.

\begin{table}[!htb]
  \centering
  \caption{PICOC framework for this systematic mapping study.}
  \label{tab:picoc}
  \small
  \begin{tabularx}{\columnwidth}{@{} l X @{}}
    \toprule
    \textbf{Element} & \textbf{Definition} \\
    \midrule
    Population      & Interactive LLM-based NPCs in games or virtual environments \\[3pt]
    Intervention    & LLM-based tools and architectures for generating realistic dialogue and autonomous behaviour, including systems with memory, retrieval grounding, planning, and tool use \\[3pt]
    Comparison      & Scripted NPC AI (behaviour trees, FSMs, GOAP, utility AI), non-LLM dialogue systems, or ablated LLM variants without memory, RAG, or planning components \\[3pt]
    Outcome         & Believability and human-likeness of NPC dialogue and behaviour; dialogue coherence and consistency; behaviour quality and agency; engineering feasibility (latency and cost); safety; and designer controllability \\[3pt]
    Context         & Interactive games and virtual worlds with real or simulated player-agent interaction \\
    \bottomrule
  \end{tabularx}
\end{table}

We defined \emph{LLM-based NPCs} broadly to include autonomous agents in simulated virtual environments (e.g., Minecraft-based research agents) where the agent fulfils a role similar to an NPC, in which they respond to player or user input, inhabiting a persistent world, and exhibiting goal-directed behaviour, even when the paper does not use the term ``NPC''. In these contexts, the end user is typically a researcher or an automated training process rather than a human player, but the underlying architectures still work for player-facing NPC design. This reflects the convergence between game LLM-based NPC research and autonomous agent research and captures architectures (e.g., Generative Agents~\cite{Park2023Generative}, DEPS~\cite{describeexplain2025}) directly applicable to NPC design despite originating outside it.

\subsection{Research Questions}
\label{sec:method:rqs}

We organised the mapping around one main research question and ten sub-questions that branch from it along four thematic branches: architecture and design, scope and evaluation, engineering and control, and quality and transparency.

The main research question of \emph{``what LLM-based approaches exist for creating lifelike NPC dialogue and behaviour, and how are they evaluated?''} is the baseline question that this review answers in Section~\ref{sec:discussion}. It branches into ten sub-questions grouped along four thematic axes:

\begin{enumerate}
    \item \textbf{Architecture \& Design}
    
    \begin{itemize}
      \item[RQ1] What LLM-based tools, architectures, and design techniques have been proposed to enable lifelike NPC dialogue and behaviour in interactive environments?
      \item[RQ2] Which system components are used to ground and control NPC outputs (e.g., memory, RAG, world-state grounding, planning, tool/function calling), and how are they combined?
    \end{itemize}
    
    \item[1.] \textbf{Scope \& Evaluation}
    
    \begin{itemize}
      \item[RQ3] What kinds of NPC capabilities are targeted (dialogue-only, behaviour-only, dialogue + action), and what interactions are supported (text, voice, embodied action)?
      \item[RQ4] In what contexts are these NPCs deployed (game genres, virtual worlds, interVR/simulators), and what assumptions do authors make about the environment and world state access?
      \item[RQ5] What baselines and comparisons are used (scripted dialogue trees, behavior trees/FSM/GOAP/utility AI, non-LLM systems, or ablations), and what improvements are reported?
    \end{itemize}

    \item \textbf{Engineering \& Control}
                                           
    \begin{itemize}
      \item[RQ6] What trade-offs are reported between realism and engineering constraints (latency, cost, reliability), and what strategies are used to meet real-time requirements?
      \item[RQ7] How is designer control handled (personality consistency, authorial intent, constraints, “lore compliance,” controllable behaviors), and what control interfaces are provided?
      \item[RQ8] What resources are released (code, prompts, datasets, benchmarks, game environments), and how reproducible are the results?
    \end{itemize}

    \item \textbf{Quality \& Transparency axis}
                                                                     
    \begin{itemize}
      \item[RQ9] How do studies address safety and alignment issues (toxicity, policy compliance, hallucinations, jailbreaks), especially in player-facing settings?
      \item[RQ10] What limitations are repeatedly reported, and what open research problems remain (evaluation validity, long-term consistency, multi-agent dynamics, personalization, etc.)?
    \end{itemize}
\end{enumerate}

\subsection{Search Strategy}
\label{sec:method:search}

We searched three digital libraries: ACM Digital Library, IEEE Xplore, and Scopus. We chose these sources because they collectively organise the most popular venues for human-computer interaction, game AI, and machine learning research, including ACM CHI, UIST, SIGGRAPH, NeurIPS, ECCV, and IEEE CoG. We set the publication date range to January 2022 to February 2026 to capture the period following the public release of instruction-tuned frontier LLMs (GPT-3.5, ChatGPT), which marks the practical starting point of LLM-based NPC research.

We made a Boolean search string using two complementary sub-queries joined by \texttt{OR}. The first sub-query targets papers that clearly combine LLM-based NPC or game concepts with LLMs and dialogue or behaviour terms. The second sub-query targets papers on generative or autonomous agents in virtual or open-world environments, which captures agent architectures applicable to LLM-based NPCs even when authors do not use the term ``NPC.'' The full search string is as shown below.

\smallskip
\noindent\texttt{((~(~``non-player character'' OR npc~) AND (~``video game'' OR videogame OR gameplay~) AND (~``large language model'' OR ``generative ai'' OR llm~) AND (~dialogue OR dialog OR behavior OR behaviour~)~)} \texttt{OR (~(~``large language model'' OR ``generative ai'' OR llm~) AND (~``generative agent'' OR ``generative agents'' OR ``autonomous agent'' OR ``autonomous agents'' OR ``embodied agent'' OR ``embodied agents'' OR ``simulated agent'' OR ``simulated agents''~) AND (~``virtual world'' OR sandbox OR ``open world''~)~))}
\smallskip

We applied the search string to the title, abstract, and keyword fields in all three libraries. The search retrieved 323 records from ACM Digital Library, 8 from IEEE Xplore, and 46 from Scopus (384 records total). We removed 7 records duplicated across databases, yielding 377 records for screening.

\subsection{Study Selection}
\label{sec:method:selection}

\paragraph{Inclusion criteria.}
We included a paper if it met both of the following conditions: (1) it focuses on the use of LLMs or generative AI models for LLM-based NPC behaviour or autonomous agent behaviour in an interactive or simulated environment; and (2) it is written in English.

\paragraph{Exclusion criteria.}
We excluded a paper if it met any of the following conditions: (1) it does not mention LLMs or AI agents in a simulated or interactive environment; (2) it does not address LLM-based NPCs or autonomous agent behaviour; or (3) it is a short abstract, poster, or workshop summary without a full technical or empirical contribution (fewer than 4 pages).

\paragraph{Selection process.}
Two reviewers independently screened all 377 records by title and abstract as detailed in Figure~\ref{fig:prisma}. From this, 327 were excluded due to their lack of conformity with the inclusion criteria as well as the presence of the exclusion criteria, resulting in 50 papers. We resolved disagreements through discussion and, when consensus was not reached, by reading the full text of the disputed record. We applied the inclusion and exclusion criteria to all records that passed title and abstract screening, reading full texts where necessary to assess eligibility. The process yielded 42 papers for full data extraction.

\subsection{Quality Assessment}
\label{sec:method:qa}

We assessed each included paper against five quality criteria derived from the PICOC outcomes (\autoref{tab:picoc}) and the specific requirements of LLM-based NPC research. Two reviewers scored each criterion independently using a three-point scale (Yes / Partially / No), and we resolved disagreements through discussion. Table~\ref{tab:qa} lists the criteria and the distribution of scores across the final 42 included papers. Each criterion was scored Yes = 1, Partial = 0.5, No = 0, and summed to a 0-5 aggregate per paper; we excluded papers that scored below 3.0. To clarify, the quality assessment phase lowered the candidate set from 50 to 42 (8 of them scored below 3.0), as shown in Figure \ref{fig:prisma}.

\begin{table}[!htb]
  \centering
  \caption{Quality assessment criteria and score distribution ($n = 42$).}
  \label{tab:qa}
  \small
  \begin{tabularx}{\columnwidth}{@{} p{5.5cm} c c c @{}}
    \toprule
    \textbf{Criterion} & \textbf{Yes} & \textbf{Partial} & \textbf{No} \\
    \midrule
    QA1: Does the paper clearly describe the LLM(s) or generative model(s) used?                                              & 23 & 11 &  8 \\[3pt]
    QA2: Is the LLM applied to NPCs or autonomous agents in an interactive or simulated environment?                           & 42 &  0 &  0 \\[3pt]
    QA3: Does the paper provide sufficient technical detail to understand or reproduce the approach?                           & 12 & 19 & 11 \\[3pt]
    QA4: Does the paper evaluate NPC behaviour or dialogue realism using qualitative or quantitative methods?                  & 35 &  5 &  2 \\[3pt]
    QA5: Is the study empirically validated (experiments, user studies, benchmarks, or simulations)?                          & 27 & 11 &  4 \\
    \bottomrule
  \end{tabularx}
\end{table}

\subsection{Data Extraction}
\label{sec:method:extraction}

 We extracted data from each included paper into a structured spreadsheet with 36 fields organised into seven thematic groups. One reviewer performed the initial extraction for each paper; a second reviewer verified all entries and flagged disagreements for discussion.

We populated categorical fields (e.g., architecture pattern, grounding components, interaction modalities) using controlled vocabularies agreed to before extraction began. We permitted multiple values per field where a paper combined more than one pattern or component. We recorded free-text evidence for each quality assessment criterion to support claims made during synthesis. For fields where we could not determine a value from the paper text alone (e.g., exact model version for papers using commercial platforms), we recorded the field as ``unclear'' and included an explanation in the evidence column rather than imputing a value. Included here is the link to the file with the described usage of each variable used for extracting data\footnote{\url{https://doi.org/10.5281/zenodo.20490290}}.

\section{Results} \label{sec:res}


We ran a systematic search across ACM Digital Library (323 candidates), IEEE Xplore (8), and Scopus (46) and obtained a final corpus of 42 papers published between 2023 and 2026 after we applied the inclusion and exclusion criteria and the quality assessment checklist.
The corpus includes conference proceedings and journals (40) and arXiv preprints articles (2).

The following subsections address each research question in turn.

\subsection*{RQ1: What LLM-based tools, architectures, and design techniques have been proposed to enable lifelike NPC dialogue and behaviour in interactive environments?}
\label{sec:rq1}

Several distinct architecture patterns appear across the corpus, ranging from single-prompt pipelines to multi-agent cognitive systems with learned components (Table~\ref{tab:architectures}).

\begin{table}[!htb]
  \centering
  \caption{Architecture patterns identified in the corpus ($n = 42$). Papers may
           be coded under more than one pattern.}
  \label{tab:architectures}
  \small
  \begin{tabularx}{\columnwidth}{@{}Xc X@{}}
    \toprule
    \textbf{Architecture Pattern} & \textbf{\# Papers} & \textbf{Paper List} \\
    \midrule
    Single-agent (prompt pipeline) & 24 &
      \cite{conversationalinteractions2024, liesdeceit2025,
            integratingenvironmental2025, llmpowerednpcs2024,
            openendednpc2025, aidrivennpcdialogues2025,orchestratingadaptive2025} \\[3pt]
    Multi-agent system & 15 &
      \cite{Park2023Generative, multiagentinteractive2025,
            traininglanguagemodels2024, metaagents2025,
            canagents2025, enhancinggenerative2025} \\[3pt]
    Planner + LLM & 17 &
      \cite{describeexplain2025, seeandthink2024, adam2025,
            classroomsimulacra2025, Sun2025TalkingToBuild,
            odyssey2024, creativeagents2025} \\[3pt]
    GOAP + LLM & 1 &
      \cite{metaagents2025} \\[3pt]
    Tool-using agent & 20 &
      \cite{whatsthegame2025, playerdriven2025, openmap2025,
            adam2025, chainofimagination2025, enhancinggame2024} \\[3pt]
    BT/FSM + LLM hybrid & 1 &
      \cite{killeronboard2025, metaagents2025} \\[3pt]
    Multi-LLM modular pipeline & 2 &
      \cite{seeandthink2024, immersivetailoring2025} \\[3pt]
    RL + LLM hybrid & 2 &
      \cite{larm2025, traininglanguagemodels2024} \\[3pt]
    Runtime code generation & 1 &
      \cite{whatsthegame2025} \\[3pt]
    Offline pre-generation & 3 &
      \cite{balancinggame2025, liesdeceit2025, killeronboard2025} \\[3pt]
    Text-conditioned simulation (diffusion) & 1 &
      \cite{textguidedsynthesis2025} \\
    \bottomrule
  \end{tabularx}
\end{table}

Researchers most often use the \textit{single-agent prompt pipeline}. In this setup, one LLM receives a composed prompt containing persona, short-term conversation history, world state, and rules, then returns a dialogue response or action description. This pattern dominates dialogue-only LLM-based NPC research \cite{conversationalinteractions2024, orchestratingadaptive2025,
integratingenvironmental2025} because developers can integrate it into commercial engines (Unity, Unreal) via API calls.


\textit{Planner + LLM} architectures break relatively extended tasks into sub-goal sequences that a separate controller or skill layer executes. DEPS \cite{describeexplain2025} introduces a four-module loop (Descriptor, Explainer, Planner, Selector) in which the LLM replans after each failed sub-goal. STEVE \cite{seeandthink2024} extends this with
four specialised sub-agents (Planner, Critic, Curriculum, Describer) and a vision encoder. ODYSSEY \cite{odyssey2024} uses a planner-actor-critic triad with a skill library. These architectures target open-world task completion rather than LLM-based NPC conversation.

STEVE and Immersive Tailoring \cite{immersivetailoring2025} exemplify the \textit{multi-LLM modular pipeline}. Immersive Tailoring decomposes LLM-based NPC behaviour into seven specialised LLM modules (Detector, Scene, Speech, Pose \& Animation, Role \& Goal, Personality, Appearance) to balance capability against per-module latency.

Killer on Board \cite{killeronboard2025} uses Unity's AI Navigation for movement and a behaviour tree for state transitions, while a local LLM (Hermes~3 3B) handles dialogue. MetaAgents \cite{metaagents2025} integrates Goal-Oriented Action Planning (GOAP) with an LLM for multi-agent job-fair simulations.

BPST \cite{balancinggame2025} uses \textit{offline pre-generation with adaptive runtime selection} and removes LLM inference from the game loop. The LLM generates multiple emotional tone variants of intent-based dialogue branches before launch, and a Pursuit Learning Automaton selects among them at runtime based on conversation-time feedback.

Across the corpus, authors most often use \textit{system prompt / persona injection} (37/42 papers), \textit{structured or JSON output} (33/42), \textit{few-shot examples} (16/42), and \textit{chain-of-thought} (12/42). About one-third of papers (12/42) use fine-tuning, primarily via LoRA adapters on open-weight models such as LLaMA-2 \cite{seeandthink2024}, LLaMA-3 \cite{multiagentinteractive2025, odyssey2024, enhancinggame2024}, and RWKV \cite{traininglanguagemodels2024}.

\subsection*{RQ2: Which system components are used to ground and control NPC outputs (e.g., memory, RAG, world-state grounding, planning, tool/function calling), and how are they combined?}
\label{sec:rq2}

Table~\ref{tab:grounding} summarises how often each grounding or control component appears in the corpus. Short-term memory and world-state injection appear in most papers, while long-term memory, reflection, and RAG appear less often. Most papers use \textit{sequential prompt concatenation} instead of learned integration: they put perception outputs into only one prompt, feed it to the LLM, and parse the response. Across the corpus, three integration patterns recur.

\begin{table}[!htb]
  \centering
  \caption{Frequency of grounding and control components ($n = 42$).}
  \label{tab:grounding}
  \small
  \begin{tabular}{@{}lcc@{}}
    \toprule
    \textbf{Component} & \textbf{\# Papers} & \textbf{~\%} \\
    \midrule
    Short-term memory (prompt history)   & 25 & 60 \\
    World-state grounding                & 29 & 69 \\
    Rules / output constraints           & 29 & 69 \\
    Emotion / personality model          & 20 & 48 \\
    Planning module                      & 23 & 55 \\
    Long-term memory (vector / summary)  & 13 & 31 \\
    Reflection / self-critique loop      & 16 & 38 \\
    Knowledge base / lore store          & 18 & 43 \\
    Tool / function calling              & 7 & 17 \\
    RAG / skill retrieval                & 6 & 14 \\
    Safety filter / guardrails           &  2 &  5 \\
    \bottomrule
  \end{tabular}
\end{table}

\paragraph*{Lightweight single-pass grounding (dialogue-only papers)} Several player-facing dialogue systems assemble a single prompt per turn from a world-state summary, persona, short-term history, and rule constraints, emitting a response in one pass with no planning module. Rule lines constrain vocabulary, enforce lore, or block out-of-character responses, as seen across architecturally simple single-agent systems\cite{integratingenvironmental2025, larm2025, towarddynamic2025}.

\paragraph*{Mid-tier looped grounding (social simulations)} Authors enhance the single-pass pattern with a long-term memory store, an emotion model that modulates next-turn behaviour, and optional reflection. Generative Agents \cite{Park2023Generative} and MetaAgents \cite{metaagents2025} are the representative examples; the Orchestrating Adaptive AI system \cite{orchestratingadaptive2025} extends this to player-facing RPG NPCs with trust vectors and event-driven memory decay, while CogniSociety \cite{enhancinggenerative2025} adds a shared vector database and emotion classification.

\paragraph*{Full cognitive loop (open-world task agents)} Agents run a repeated perception cycle: memory retrieval, reflection, planning, execution, and memory update cycle, with each step conditioning the next. ADAM \cite{adam2025}, STEVE \cite{seeandthink2024}, and ODYSSEY \cite{odyssey2024} exemplify this pattern; LARM \cite{larm2025} replaces prompt chaining with end-to-end RL via PPO, and GROMIT \cite{whatsthegame2025} inverts it entirely by operating statelessly, reconstructing a fresh Semantic Scene Graph on every invocation to avoid memory drift.

\subsection*{RQ3: What kinds of NPC capabilities are targeted (dialogue-only, behaviour-only, dialogue + action), and what interactions are supported (text, voice, embodied action)?}
\label{sec:rq3}

\begin{figure}[h]
  \centering
  \begin{tikzpicture}
    \pie[
      radius=1.5,
      sum=42,
    after number=,   
    text=legend,
    color={
      blue!40,
      red!40,
      green!40,
      orange!40,
  },
    ]{
      15/Dialogue-Only,
      15/Dialogue and Action Oriented,
      10/Behavior-Only,
      2/Unclear
    }
  \end{tikzpicture}
  \caption{The methods of behavior the LLM-based NPCs/Agents were programmed to perform.}
  \label{fig:npc-behavior-breakdown}
\end{figure}
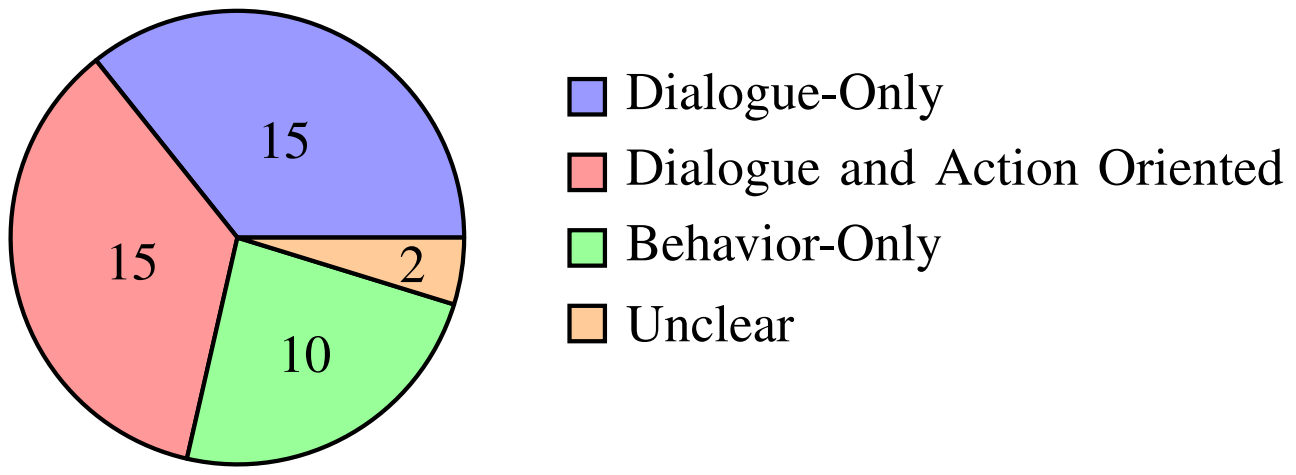

The corpus is divided into three groups (Figure~\ref{fig:npc-behavior-breakdown}). \textit{Dialogue-only} systems (15/42) restrict the LLM to generating speech and do not give it control over agent movement or game state \cite{conversationalinteractions2024, liesdeceit2025, balancinggame2025, llmpowerednpcs2024, openendednpc2025}. \textit{Dialogue and action integrated} systems (15/42) let the LLM emit speech and action commands, and the game engine parses and executes those commands \cite{Park2023Generative, playerdriven2025, Sun2025TalkingToBuild, immersivetailoring2025, killeronboard2025, bartendersadvice2025, metaagents2025}. \textit{Behaviour-only} systems (10/42) target autonomous task completion without conversational dialogue as the primary objective \cite{whatsthegame2025, describeexplain2025, seeandthink2024, adam2025, textguidedsynthesis2025, classroomsimulacra2025, larm2025, odyssey2024}.

\textit{Text} input and output appear 36 times across the corpus. \textit{Voice} (speech-to-text and text-to-speech) appears in eight papers \cite{conversationalinteractions2024, immersivetailoring2025, myfocus2025, unlockingthepower2024, socialpresence2025, canyoufeel2024}, primarily in VR or accessibility-oriented contexts. \textit{Embodied action} (the LLM-based NPC's physical presence or movement in a 3D world) appears in 19 papers, almost exclusively in Minecraft-based research or VR \cite{describeexplain2025, seeandthink2024, adam2025, larm2025, odyssey2024, metaagents2025, immersivetailoring2025, socialpresence2025}. \textit{Multimodal perception} (processing visual game frames) appears in 10 papers \cite{seeandthink2024, adam2025, larm2025, chainofimagination2025, immersivetailoring2025, canai2024} and remains largely confined to research agents rather than deployed LLM-based NPC systems. \textit{Gesture and animation} output appears rarely (4 papers: \cite{Sun2025MyselfMyChildMyFriend, socialpresence2025, immersivetailoring2025, towarddynamic2025}), and a separate animation system drives it rather than the LLM.

\subsection*{RQ4: In what contexts are these NPCs deployed (game genres, virtual worlds, interVR/simulators), and what assumptions do authors make about the environment and world state access?}
\label{sec:rq4}

The corpus spans six deployment contexts. \textit{Open-world survival games}, primarily Minecraft, dominate the behaviour-only sub-corpus (DEPS \cite{describeexplain2025}, STEVE \cite{seeandthink2024}, ADAM \cite{adam2025}, LARM \cite{larm2025}, ODYSSEY \cite{odyssey2024}, Talking-to-Build \cite{Sun2025TalkingToBuild}, Creative Agents \cite{creativeagents2025}). \textit{RPG and interactive fiction} settings dominate LLM-based NPC dialogue work, spanning commercial titles via modding (Skyrim + Mantella \cite{llmpowerednpcs2024}), custom mystery games (Vaudeville \cite{conversationalinteractions2024}, Killer on Board \cite{killeronboard2025}, Dejaboom! \cite{playerdriven2025}), and purpose-built prototypes \cite{balancinggame2025, openendednpc2025, orchestratingadaptive2025}. \textit{Social simulations} form a growing category \cite{Park2023Generative, metaagents2025, canagents2025, enhancinggenerative2025}. \textit{Serious games and training simulators} target educational or therapeutic goals \cite{multiagentinteractive2025, classroomsimulacra2025, myfocus2025, unlockingthepower2024, bartendersadvice2025}. \textit{VR / XR environments} appear in three papers \cite{myfocus2025, socialpresence2025, immersivetailoring2025}. Three papers \cite{classroomsimulacra2025, anllm2025, openmap2025} apply their architectures to non-game simulation domains (education, energy, indoor navigation).

Thirty-three papers assume \emph{partial} world-state access and inject a structured text summary of inventory, scene objects, or LLM-based NPC vicinity into the prompt. Only five papers grant the LLM \emph{full} scene access via a live scene graph \cite{whatsthegame2025, textguidedsynthesis2025, orchestratingadaptive2025}. Four papers assume \emph{no formal world-state access} and rely on conversational history and persona \cite{liesdeceit2025, myfocus2025, balancinggame2025}. Eleven papers operate in \emph{simulated-only} settings with no interactive game environment \cite{anllm2025, traininglanguagemodels2024, enhancinggenerative2025, Park2023Generative}. 

Player-facing dialogue systems reduce the game world to flat object inventories, discarding geometric relationships that open-world task agents partially recover through visual token encoders \cite{seeandthink2024} and multimodal goal representations \cite{odyssey2024}.

This distinction supports the need for a sensitivity analysis comparing the two subsets across architecture, latency, safety, and reproducibility (Section~\ref{sec:disc:threats}).

\subsection*{RQ5: What baselines and comparisons are used (scripted dialogue trees, behavior trees/FSM/GOAP/utility AI, non-LLM systems, or ablations), and what improvements are reported?}
\label{sec:rq5}

Authors most often compare against ablations (19/42 papers), followed by non-LLM ML baselines such as behaviour-cloning policies or classical RL (9/42), scripted dialogue trees or static NPC conditions (7/42), and human-expert baselines (5/42). Notably, 16 papers include \emph{no baseline} of any kind \cite{integratingenvironmental2025, liesdeceit2025, whatsthegame2025, larm2025, canai2024, anllm2025} and rely exclusively on descriptive or qualitative evaluation.

Where baselines exist, authors report gains. For example, Generative Agents \cite{Park2023Generative} demonstrates via ablation that the full architecture exhibits the most believable behaviour, with agent conduct rated more believable than both reduced variants and human crowdworker descriptions. ADAM \cite{adam2025} reaches diamond acquisition in 2.2$\times$ fewer exploration steps than the prior SOTA agent and achieves a 4.6$\times$ speedup on iron acquisition in a modified environment. The Classroom Simulacra TIR module \cite{classroomsimulacra2025} yields statistically significant improvement via paired t-test over baseline performance. For player-facing LLM-based NPC dialogue, Killer on Board \cite{killeronboard2025} reports a GUESS composite score of 40.21 with a local 3B model with procedurally generated NPC identities in a Unity IDS prototype. These results should be noted as reflecting individual study conditions and have not been independently replicated.

\subsection*{RQ6: What trade-offs are reported between realism and engineering constraints (latency, cost, reliability), and what strategies are used to meet real-time requirements?}
\label{sec:rq6}

32 of 42 papers treat real-time feasibility as a concern, but only seven report numerical latency measurements (Table~\ref{tab:latency}). End-to-end response times for cloud-hosted frontier models (GPT-4, GPT-4o) range from roughly 2.6\,s to 15\,s, well above the sub-second threshold that game UX literature commonly cites. Player-Driven Emergence \cite{playerdriven2025} reports 15\,s as detrimental to player experience, and 40\% of participants identify latency as a pain point. The Immersive Tailoring system \cite{immersivetailoring2025} reports 7.3\,s for a tailoring interaction and 5.2\,s for a standard interaction despite using GPT-4o-mini, and it attributes the dominant bottleneck to the Scene LLM's visual processing step (3.2\,s mean).

\begin{table}[!htb]
  \centering
  \caption{Latency measurements reported in the corpus. Papers that acknowledge
           latency qualitatively, but provide no figures are excluded.}
  \label{tab:latency}
  \small
  \begin{tabular}{@{}llr@{}}
    \toprule
    \textbf{Paper} & \textbf{Condition} & \textbf{Latency} \\
    \midrule
    GROMIT~\cite{whatsthegame2025}
      & GPT-4, Sandbox        & 10{,}600\,ms \\
    & GPT-4o, Sandbox       &  5{,}200\,ms \\
    & GPT-4, Escape Room    &  5{,}600\,ms \\
    & GPT-4o, Escape Room   &  2{,}600\,ms \\
    Player-Driven Emergence~\cite{playerdriven2025}
      & GPT-4 (game loop)     & 15{,}000\,ms \\
    Immersive Tailoring~\cite{immersivetailoring2025}
      & Tailoring path        &  7{,}300\,ms \\
    & Non-tailoring path    &  5{,}200\,ms \\
    Enhancing Game AI~\cite{enhancinggame2024}
      & BTree generation      & 147{,}000\,ms \\
    Text-Crowd~\cite{textguidedsynthesis2025}
      & Map generation        & $<$6{,}000\,ms \\
    & Runtime (per step)    & $<$1\,ms \\
    LARM~\cite{larm2025}
      & Inference (embodied)  &     580\,ms \\
    BPST~\cite{balancinggame2025}
      & Runtime selection     & $<$1{,}000\,ms \\
    \bottomrule
  \end{tabular}
\end{table}

Authors use several strategies to meet real-time or resource constraints. The most prevalent is switching to smaller or lighter models (14/42): replacing GPT-4 with GPT-4o, using LLaMA or RWKV variants, or applying LoRA fine-tuning \cite{multiagentinteractive2025, immersivetailoring2025, killeronboard2025, enhancinggame2024, larm2025}. Local model deployment (4/42) eliminates API round-trip latency entirely \cite{killeronboard2025, multiagentinteractive2025, balancinggame2025}, and fallback scripted behaviour (4/42) defaults to a static response or idle animation on timeout \cite{conversationalinteractions2024, whatsthegame2025, towarddynamic2025}. Offline pre-generation (3/42) removes LLM calls from the game loop altogether \cite{balancinggame2025, npcdialogue2024, killeronboard2025}. The remaining strategies, such as structured output with retry loops, parallel modular processing, retrieval prefetch and caching, and memory compression, each appear in 2–4 papers  \cite{describeexplain2025, adam2025, seeandthink2024, playerdriven2025, immersivetailoring2025, pangea2025} and address invalid generation, critical-path latency, repeated inference, and context window limits, respectively.

Several papers delve into cost constraints qualitatively, but none quantify them. Authors repeatedly flag dependence on closed APIs (GPT-4/Codex) as a scalability barrier and position local fine-tuned models and offline pre-generation as additional mitigations.

\subsection*{RQ7: How is designer control handled (personality consistency, authorial intent, constraints, “lore compliance,” controllable behaviors), and what control interfaces are provided?}
\label{sec:rq7}

Table~\ref{tab:control} summarizes the designer control mechanisms identified in the corpus. Authors use personality profiles and prompt templates in most papers, while authoring tools that provide a UI abstraction for non-technical designers remain rare.

\begin{table}[!htb]
  \centering
  \caption{Designer control mechanisms and their frequency ($n = 42$).}
  \label{tab:control}
  \small
  \begin{tabular}{@{}lcc@{}}
    \toprule
    \textbf{Control Mechanism}          & \textbf{\# Papers} & \textbf{\%} \\
    \midrule
    Personality profiles / persona      & 26 & 62 \\
    Prompt templates                    & 38 & 90 \\
    Output constraints / rules          & 27 & 64 \\
    Controllable parameters (sliders,   & 25 & 60 \\
    ~~~weights, budgets)                &    &    \\
    Lore knowledge base                 & 11 & 26 \\
    Authoring tool / designer UI        &  5 & 12 \\
    Behaviour budgets (token caps,      &  3 & 7 \\
    ~~~action limits)                   &    &    \\
    \bottomrule
  \end{tabular}
\end{table}

Several papers develop notable control approaches beyond generic persona prompting. GROMIT \cite{whatsthegame2025} introduces \emph{SSG filtering} as a designer control lever: designers scope the Semantic Scene Graph passed to the LLM at design time to prevent generated code from accessing unintended scene elements. BPST \cite{balancinggame2025} inverts the control problem by making the LLM generate a finite set of pre-approved dialogue variants offline and leaving runtime selection to a deterministic algorithm with no LLM agency. Immersive Tailoring \cite{immersivetailoring2025} has a \emph{player-facing} tailoring interface where users modify LLM-based NPC personality, appearance, and goals in natural language at runtime, giving players leverage over the designers. Killer on Board \cite{killeronboard2025} gives authors direct control over who the LLM-based NPC is, its relationships, schedules, observations, and ``protective scenes'' from a fixed template.

Authors often treat prompt templates as sufficient for control, yet these templates are weaker against adversarial input and do not work in complex game worlds.

\subsection*{RQ8: What resources are released (code, prompts, datasets, benchmarks, game environments), and how reproducible are the results?}
\label{sec:rq8}

As shown in Table~\ref{tab:artifacts}, fewer than one-third of papers release source code. Papers releasing full repositories include Generative Agents \cite{Park2023Generative}, OpenMap \cite{openmap2025}, DEPS \cite{describeexplain2025}, Text-Crowd \cite{textguidedsynthesis2025}, ADAM \cite{adam2025}, Classroom Simulacra \cite{classroomsimulacra2025}, Training Language Models \cite{traininglanguagemodels2024}, Killer on Board \cite{killeronboard2025}, Enhancing Game AI \cite{enhancinggame2024}, the Effect of LLM-Based NPC Emotional States \cite{theeffect2025}, Player-Driven Emergence \cite{playerdriven2025}, Creative Agents \cite{creativeagents2025}, and ODYSSEY \cite{odyssey2024}.
Five categories of reproducibility barrier recur across the corpus:

\begin{table}[!htb]
  \centering
  \caption{Artifact availability across the corpus ($n = 42$).}
  \label{tab:artifacts}
  \small
  \begin{tabular}{@{}lcc@{}}
    \toprule
    \textbf{Artifact Type} & \textbf{\# Papers} & \textbf{\%} \\
    \midrule
    Source code / repository   & 13 & 31 \\
    Prompts (full text)        & 10 & 24 \\
    Dataset                    & 13 & 31 \\
    Benchmark / environment    &  6 & 14 \\
    Demo video                 & 10 & 24 \\
    Nothing released           & 18 & 43 \\
    \bottomrule
  \end{tabular}
\end{table}

Deprecated model APIs that prevent exact replication \cite{describeexplain2025, adam2025, seeandthink2024}; withheld system prompts and few-shot templates (32 papers) \cite{whatsthegame2025, integratingenvironmental2025, Sun2025TalkingToBuild, anllm2025, bartendersadvice2025, pangea2025}; commercial platform opacity (6 papers) from Inworld AI, Convai, or Worqhat \cite{conversationalinteractions2024, theeffect2025, socialpresence2025, canyoufeel2024, orchestratingadaptive2025}; unnamed or underspecified LLMs (8 papers) \cite{balancinggame2025, openendednpc2025, bartendersadvice2025, enhancinggenerative2025, canagents2025}; and paywalled game or dataset requirements such as Minecraft's commercial licence and Gameloft's unreleased behaviour-tree datasets \cite{enhancinggame2024}. Fewer than 30\% of papers have sufficient transparency for independent replication, a systemic reproducibility gap in the field.

\subsection*{RQ9: How do studies address safety and alignment issues (toxicity, policy compliance, hallucinations, jailbreaks), especially in player-facing settings?}
\label{sec:rq?}

Authors report hallucination as the most critical safety concern (13/42), followed by bias (11)\cite{liesdeceit2025, anllm2025, metaagents2025, odyssey2024, Sun2025TalkingToBuild, Sun2025MyselfMyChildMyFriend, immersivetailoring2025, canagents2025, enhancinggenerative2025, pangea2025, conversationalinteractions2024}, toxicity (4)\cite{Sun2025TalkingToBuild, metaagents2025, canagents2025, bartendersadvice2025}, privacy (5)\cite{classroomsimulacra2025, anllm2025, immersivetailoring2025, odyssey2024, canyoufeel2024}, player jailbreaking (2)\cite{conversationalinteractions2024, whatsthegame2025}, and therapeutic safety (1)\cite{bartendersadvice2025}. The most common remedy is persona and role constraint via system prompt (37/42), which is cheap to implement but vulnerable to adversarial input. Stronger measures are rare: constrained decoding appears in two papers \cite{adam2025, killeronboard2025}, human-in-the-loop design review in two others \cite{liesdeceit2025, bartendersadvice2025}, and a moderation API filter in one \cite{Sun2025TalkingToBuild}. Only two papers implement runtime guardrail components as actual system elements \cite{multiagentinteractive2025, killeronboard2025}; the rest either treat safety as a known limitation or depend only on prompt engineering.

\subsection*{RQ10: What limitations are repeatedly reported, and what open research problems remain (evaluation validity, long-term consistency, multi-agent dynamics, personalization, etc.)?}
\label{sec:rq10}

Table~\ref{tab:openproblems} lists the ten most frequently cited open problems, and hallucination, latency, and evaluation validity form a triad that appears in the limitations sections of most papers.

\begin{table}[!htb]
  \centering
  \caption{Open problems identified by authors across the corpus, ranked by frequency
           of mention ($n = 42$).}
  \label{tab:openproblems}
  \small
  \begin{tabular}{@{}lc@{}}
    \toprule
    \textbf{Open Problem} & \textbf{\# Papers} \\
    \midrule
    Evaluation validity (no benchmark, small $n$)     & 33 \\
    Hallucination and factual inconsistency           & 26 \\
    Real-time latency                                 & 24 \\
    Long-term memory consistency                      & 19 \\
    Controllability vs.\ emergent behaviour           & 23 \\
    Dataset and benchmark gaps                        & 16 \\
    Personalization across sessions                   & 22 \\
    Safety in player-facing deployment                & 8 \\
    Multi-agent dynamics and inter-NPC coordination   & 10 \\
    Memory scaling to large agent populations         & 8 \\
    \bottomrule
  \end{tabular}
\end{table}

\paragraph*{Evaluation validity.} Researchers lack a shared LLM-based NPC believability benchmark, and papers use incompatible scales, including: GUESS-18 \cite{killeronboard2025}, PXI Mini \cite{canyoufeel2024}, GEQ \cite{Sun2025TalkingToBuild}, SUS \cite{Sun2025MyselfMyChildMyFriend}, MSE, NMMSP \cite{socialpresence2025}, success rate \cite{describeexplain2025}, BLEU/ROUGE \cite{multiagentinteractive2025}; this prevents direct comparison across studies. User studies typically draw small, demographically narrow samples (medians of $n \approx 25$--30, often university students in a single locale), which limits generalisation.

\paragraph*{Long-term consistency.} LLM-based NPCs routinely contradict themselves within a session and fail to carry narrative state across sessions, despite attempts using memory compression \cite{Park2023Generative, seeandthink2024}, dual memory architectures\cite{orchestratingadaptive2025, pangea2025, bartendersadvice2025}, and retrieval-augmented grounding \cite{seeandthink2024, enhancinggame2024}, but no solution achieves sustained cross-session coherence.

\paragraph*{Controllability versus emergence.} Scripted approaches guarantee authorial consistency but eliminate emergent behaviour, while fully open LLMs produce novel interactions but remain unpredictable; BPST \cite{balancinggame2025}, and Killer on Board \cite{killeronboard2025} offer partial mitigations, but neither provides a general-purpose resolution. The Open-Ended NPC Dialogue study \cite{openendednpc2025} shows the optimal design point depends on player type.

\paragraph*{Multi-agent dynamics.} Most reviewed architectures do not let LLM-based NPCs track other NPCs' states or dialogue history, with Generative Agents \cite{Park2023Generative} and MetaAgents \cite{metaagents2025} being standouts here. Social emergence in multi-NPC systems (information propagation, coalition formation, emergent narrative conflict) remains largely unexplored in player-facing settings.

\paragraph*{Personalization.} Robust persona prompts are the dominant individuation tactic, but do not adapt to player behaviour across sessions; BPST's Pursuit Learning Automaton \cite{balancinggame2025} is the only reviewed system with dynamic player-feedback-driven adaptation, and it operates only over pre-generated dialogue. Twenty-two papers identify cross-session adaptive personality modelling as an open problem, and no reviewed system addresses it.

\paragraph*{Safety infrastructure.} Hallucination is the most frequently reported reliability issue, yet only two papers implement runtime guardrail components \cite{multiagentinteractive2025, killeronboard2025}. Thirteen of 42 papers report hallucination as the primary safety concern, while a separate 26 identify it as an open research problem(Table~\ref{tab:openproblems}). Deployments in child-facing \cite{multiagentinteractive2025} or therapeutic contexts \cite{bartendersadvice2025} raise the stakes, yet neither has been stress-tested against adversarial input. 

\section{Discussion} \label{sec:discussion}

\subsection{Research Gaps and Open Directions}
\label{sec:disc:gaps}

\paragraph*{Evaluation infrastructure}
The most noticeable gap is the absence of a unified benchmark for LLM-based NPC believability. A purpose-built benchmark could include (1) a publicly accessible game environment, (2) a fixed set of player-NPC scenarios with annotated ground-truth lore/dialogues, and (3) a standardised player-study protocol with pre-registered hypotheses.

\paragraph*{Long-term and cross-session memory}
Every paper that attempts cross-turn memory acknowledges that context windows fill, summarisation loses detail, and LLM-based NPCs eventually contradict themselves.
Therefore, the main bottleneck for believable LLM-based NPCs may no longer be dialogue generation itself, but memory persistence and identity maintenance over time. 
One step towards this is the Classroom Simulacra's Transferable Iterative Reflection mechanism~\cite{classroomsimulacra2025}, which generalises learned behavioural patterns across new groups. A related limitation is spatial encoding: flat text lists of objects are main world-state format, without geometric relationships that vision-language pipelines in STEVE~\cite{seeandthink2024} and Chain-of-Imagination~\cite{chainofimagination2025} show are recoverable in game-like projects, but neither has conversational NPCs.

\paragraph*{Multi-agent social dynamics}
Only two reviewed architectures have NPCs communicating with each other in a game or game-like context~\cite{Park2023Generative,metaagents2025}, yet emergent NPC-to-NPC behaviour is what open-world players expect. Extending social deduction work~\cite{traininglanguagemodels2024} to narrative RPG contexts (e.g., gossip, alliances, and player-driven plot) remains unexplored.

\paragraph*{Player-adaptive personalisation}
BPST~\cite{balancinggame2025} shows that an LLM-based NPC dialogue tone can shift over time, even operating on a fixed vocabulary of pre-generated types. Genuine personalisation, like updating conversational style and relationship stance from collected session data, is technically within reach of retrieval-augmented architectures, but no reviewed paper attempts it. 

\paragraph*{Comparing LLM Implementation Results}
Comparing results across LLM-based NPC implementations is difficult because the corpus uses incompatible evaluation instruments (success rates, BLEU/ROUGE scores, GUESS-18, PXI Mini, GEQ, SUS, MSE, and NMMSP) often for single-study designs with no shared baseline. A system with 85\% task success and one reporting a mean GEQ score of 3.8 cannot be meaningfully ranked, even when targeting the same NPC capability. Until a purpose-built believability benchmark exists, unified reporting standards (end-to-end latency, sample size, and at least one validated player-experience scale with task metrics) would let future reviews make cross-study comparisons that this one cannot.

\subsection{Barriers to Commercial Adoption}
\label{sec:disc:commercial}

\subsubsection*{[Barrier\#1] Latency.}
The corpus treats latency as an engineering inconvenience, but not as a disqualifying issue. \emph{Papers reporting per-turn response times of 7, 10, and 15 seconds frame these figures as manageable, but response times of this magnitude could be incompatible with real-time commercial game genres, though tolerance could vary substantially by genre and interaction design}. Of all the noticed patterns, latency is the biggest practical hurdle in the corpus.

Most user studies run under conditions tolerant of delay: short sessions, turn-based or text-only interaction, and no competitive pressure.

The papers' suggested fix, substituting smaller models, has a quality tradeoff that the corpus acknowledges but does not systematically measure. Switching to smaller or lighter models is the most common latency mitigation in the corpus, adopted by 14 of 42 papers, yet this substitution introduces a quality tradeoff that no reviewed paper measures directly. Killer on Board~\cite{killeronboard2025} explicitly acknowledges that its locally-deployed 3B model proved insufficient for distinct NPC personality and role-playing quality relative to larger alternatives. STEVE similarly documents that its fine-tuned 13B variant required more task iterations than its GPT-4 counterpart~\cite{seeandthink2024}, and LARM reports performance degradation when smaller base models replace larger ones~\cite{larm2025}. These cases suggest that \emph{reducing model size might lower latency at the cost of the consistency and believability properties that make LLM-based NPCs compelling} in the first place. Researchers and developers could benefit from studies that treat this as a controlled variable, measuring, for a fixed NPC task, how response quality, hallucination rate, and character consistency degrade as model size decreases, so that the latency-quality frontier can be mapped rather than assumed.

\subsubsection*{[Barrier\#2] Consistency and quality assurance.}
Commercial games need deterministic content review, but LLM outputs are non-deterministic. Even at temperature zero, minor context changes can alter outputs. Toward Dynamic NPC~\cite{towarddynamic2025} reports that short personality prompts ``can't cover all scenarios'' and that GPT fills gaps with unintended violence.

\subsubsection*{[Barrier\#3] Cost at player scale.}
A title with one million daily active players generating 50 LLM-based NPC exchanges each produces 50 million API calls per day. This suggests a potentially significant financial burden for studios operating at consumer scale, though actual costs would depend heavily on caching strategies, model selection, and dialogue frequency. No reviewed paper quantifies cost at player scale; the previous estimate is the authors' own calculation based on publicly available API pricing, intended to illustrate the scope of the problem. Running the model locally avoids API costs but means either bundling large model files with the game or maintaining your own server.

\subsubsection*{[Barrier\#4] Multiplayer synchronisation.}
All reviewed architectures assume a single player interacting with an LLM-based NPC; Section~\ref{sec:disc:technical} discusses the synchronisation, consistency, and security challenges that arise in multiplayer settings.

\subsubsection*{[Barrier\#5] Content rating and regulatory compliance.}
LLMs can say almost anything, which makes it impossible for studios to guarantee a content rating. Players in the Vaudeville study~\cite{conversationalinteractions2024} actively jailbroke NPCs and shared working exploits with each other; GROMIT acknowledges the same problem~\cite{whatsthegame2025} but leaves the fix to future work.

\subsubsection*{[Barrier\#6] Localisation.}
Frontier LLMs work noticeably worse in non-English languages, and the gap widens for languages underrepresented in training data. Fine-tuning smaller models for specific languages helps. The Japanese escape game~\cite{npcdialogue2024} study, for example, identifies ELYZA and rinna as candidates. The issue here, though, is that each language needs its own pipeline, which multiplies cost and the chance of culturally tone-deaf outputs.

\subsection{Copyright, Security, and the Impact on Creative Labour}
\label{sec:disc:copyright}

The corpus does not address whether LLM-based NPC dialogue reproduces copyrighted material, leaving ownership of generated story beats an open question. Fine-tuning on domain-specific  data~\cite{multiagentinteractive2025,killeronboard2025} partially mitigates this but cannot eliminate the risk when base model training data remains undisclosed. Two papers document active jailbreaking in player-facing deployments~\cite{conversationalinteractions2024,whatsthegame2025}, and prompt injection via in-game text represents an additional unresolved attack surface, yet no reviewed paper brings a security-oriented workaround beyond prompt-level defences. 

Beyond security, the corpus neither addresses the privacy obligations (GDPR, COPPA) arising from routing player dialogue through commercial APIs, nor measures whether the labour shift from writing dialogue to engineering prompts~\cite{Park2023Generative,killeronboard2025,
pangea2025} actually reduces authoring cost, and several systems pair LLM output with text-to-speech~\cite{immersivetailoring2025,
socialpresence2025,canyoufeel2024}, leaving the legal and ethical stakes of synthesised performance unexamined.

\subsection{Technical Roadblocks on the Software Side}
\label{sec:disc:technical}


\subsubsection*{[S.E. Challenge \#1] Game engine integration.}
Unity and Unreal were built for deterministic game logic, not asynchronous API calls with multi-second response times, so integrating an LLM requires either stalling the main thread or building custom async systems to keep animations and NPC logic running while the model processes. As of the reviewed period, no standardised LLM integration layer had emerged in mainstream engines, and third-party packages like LLMUnity~\cite{killeronboard2025} only partially fill the gap.

\subsubsection*{[S.E. Challenge \#2] Determinism and save-state reproducibility.}
A game cannot trivially save and restore runtime-generated LLM-based NPC dialogue: memory, conversation history, and model random seed would all need serialisation; critical for branching narratives where LLM-based NPC dialogue has downstream consequences.

\subsubsection*{[S.E. Challenge \#3] Testing and quality assurance methodology.}
Standard quality assurance does not work on outputs that are different every time, and none addresses what happens when the underlying model gets updated or retired; a real concern given that DEPS~\cite{describeexplain2025} and ADAM~\cite{adam2025} rely on model versions no longer available.

\subsubsection*{[S.E. Challenge \#4] Context window limits as a game-state bottleneck.}
An open-world NPC realistically needs faction standings, active quests, and hundreds of world-state flags; far more than fits in a context window, and RAG only helps if the retrieval step already knows what the player is trying to do, which is precisely what the LLM is being asked to figure out.

\subsubsection*{[S.E. Challenge \#5] Multiplayer and shared-world state synchronisation.}
Multiplayer LLM-based NPCs introduce a synchronisation problem (\ref{sec:disc:commercial}) and an additional attack surface, where adversarial players can influence an NPC's state in ways that affect all subsequent players in that session.

None of these five challenges has a reviewed solution; closing them is a prerequisite for commercial deployment.

\begin{highlight}
Longer-term adoption of LLM-based NPCs hinges on three developments: (1) cost and speed improvements for sub-2\,B parameter local models; (2) legal clarity on generated-content ownership and training-data licensing; and (3) a shared evaluation benchmark. None requires a research breakthrough, but none is imminent.
\end{highlight}

\subsection{Threats to Validity}
\label{sec:disc:threats}

Extraction subjectivity is the primary internal threat. Inconsistent terminology across papers required interpretive judgment. For example, distinguishing a planner+LLM architecture from an augmented single-agent system is not always clear-cut. We mitigated this with a fixed coding schema applied consistently across all 42 papers.

Comparing player-facing NPC studies (20/42) against broader simulation/agent studies (22/42) for architecture, latency, safety, and reproducibility, we get two groups. For clarity, this was introduced in the Results for RQ4. Player-facing papers prefer single-agent prompt pipelines (17/20, 85\%) over the multi-agent and planner+LLM designs, which are more common in broader-agent work (11/22, 50\% and 13/22, 59\%). Broader-agent papers release code and artifacts more often (code: 10/22, 45\% vs. 3/20, 15\%; any artifact: 15/22, 68\% vs. 9/20, 45\%). Real-time constraints and safety coverage are comparable across subsets (85\% vs 68\%; 55\% vs 50\%), showing these are overall corpus patterns. These differences don't weaken the review's findings. The reproducibility gap (RQ8), the lack of safety guardrails (RQ9), and the latency-quality compromise (\ref{sec:disc:commercial}) hold within both subsets. We still have the broader corpus from this, with "LLM-based NPC and adjacent virtual-agent systems" where a claim involves both groups. 

The corpus covers only three databases (ACM Digital Library, IEEE Xplore, Scopus) and papers published between 2023 and early 2026. Industry technical reports, workshop papers without proceedings, and unindexed preprints are systematically absent.

Rapid technological change threatens this review's durability. Findings tied to specific model generations (GPT-4, GPT-4o, 3--7\,B local models) may date quickly; readers should treat quantitative results as indicative of the 2023--2026 period.


\section{Conclusion}
\label{sec:conclusion}

NPCs have always been bound by what designers explicitly script. LLMs introduced the possibility of characters that converse openly, adapt to unforeseen situations, and exhibit behaviours not hand-authored. This review systematically examines whether the field has realised that possibility, for whom, and under what conditions. Across 42 papers published between 2023 and 2026, the findings reveal substantial architectural diversity alongside a critical failure to build shared infrastructure, with reproducibility, evaluation validity, and runtime safety all remaining structurally untouched.

Future work should prioritise the following: First, the field would benefit from a shared benchmark so researchers can compare believability and coherence claims directly. Second, long-term cross-session memory remains the most critical unsolved problem. Third, hybrid designs offer the clearest path to commercial viability, yet only two papers in this corpus test them. Addressing these gaps, alongside the falling cost and latency of smaller local models, could close the distance between research prototypes and shipped games.

\clearpage

\bibliographystyle{IEEEtran}
\bibliography{main}

\end{document}